\documentclass[12pt]{elsart}
\usepackage{epsfig}
\usepackage{times}
\newcommand{\Mark}{}
\newcommand{\Markline}[2]{\marginpar{\rule[#1]{0mm}{#2}}}

\journal{{\tt Astroparticle Physics}}

\runauthor{T.~Antoni et al. (KASCADE Collaboration) }
\runtitle{Measurements of Attenuation and Absorption Lengths}

\begin{document}
\begin{frontmatter}

\title{Measurements of Attenuation and Absorption Lengths with 
the KASCADE Experiment}

\hyphenation{CRES KRETA KASCADE GHEISHA Karls-ruhe}

\author[KA-Uni]{T.~Antoni},
\author[KA-FZK]{W.\,D.~Apel},
\author[KA-Uni]{A.F.~Badea\thanksref{loa}}\relax
\thanks[loa]{on leave of absence from NIPNE, Bucharest},
\author[KA-FZK]{K.~Bekk},
\author[KA-FZK]{A.~Bercuci\thanksref{loa}},
\author[KA-FZK,KA-Uni]{H.~Bl\"umer},
\author[KA-FZK]{H.~Bozdog},
\author[BU]{I.\,M.~Brancus},
\author[KA-Uni]{C.~B\"uttner},
\author[YE]{A.~Chilingarian},
\author[KA-Uni]{K.~Daumiller},
\author[KA-FZK]{P.~Doll},
\author[KA-FZK]{R.~Engel},
\author[KA-FZK]{J.~Engler},
\author[KA-FZK]{F.~Fe{\ss}ler},
\author[KA-FZK]{H.\,J.~Gils},
\author[KA-Uni]{R.~Glasstetter},
\author[KA-Uni]{R.~Haeusler},
\author[KA-FZK]{A.~Haungs},
\author[KA-FZK]{D.~Heck},
\author[KA-Uni]{J.\,R.~H\"orandel},
\author[KA-Uni]{A.~Iwan\thanksref{lodz}}\relax
\thanks[lodz]{and University of Lodz, Lodz, Poland},
\author[KA-Uni,KA-FZK]{K.-H.~Kampert},
\author[KA-FZK]{H.\,O.~Klages},
\author[KA-FZK]{G.\,Maier\thanksref{author}}\relax
\thanks[author]{corresponding author, email: gernot.maier@ik.fzk.de},
\author[KA-FZK]{H.\,J.~Mathes},
\author[KA-FZK]{H.\,J.~Mayer},
\author[KA-FZK]{J.~Milke},
\author[KA-FZK]{M.~M\"uller},
\author[KA-FZK]{R.~Obenland},
\author[KA-FZK]{J.~Oehlschl\"ager},
\author[KA-Uni]{S.~Ostapchenko\thanksref{los}}\relax
\thanks[los]{on leave of absence from Moscow State University, Moscow, Russia},
\author[BU]{M.~Petcu},
\author[KA-FZK]{H.~Rebel},
\author[KA-FZK]{M.~Risse},
\author[KA-FZK]{M.~Roth},
\author[KA-FZK]{G.~Schatz},
\author[KA-FZK]{H.~Schieler},
\author[KA-FZK]{J.~Scholz},
\author[KA-FZK]{T.~Thouw},
\author[KA-FZK]{H.~Ulrich},
\author[YE]{A.~Vardanyan},
\author[KA-Uni]{J.\,H.~Weber},
\author[KA-FZK]{A.~Weindl},
\author[KA-FZK]{J.~Wentz},
\author[KA-FZK]{J.~Wochele},
\author[LZ-Sol]{J.~Zabierowski}

\collab{(The KASCADE Collaboration)}

\address[KA-Uni]{Institut f\"ur Experimentelle Kernphysik, University of
             Karlsruhe, 76021~Karlsruhe, Germany}
\address[KA-FZK]{Institut f\"ur Kernphysik, Forschungszentrum Karlsruhe,
      	     76021~Karlsruhe, Germany}
\address[BU]{National Institute of Physics and Nuclear Engineering, 
             7690~Bucharest, Romania}
\address[YE]{Cosmic Ray Division, Yerevan Physics Institute, 
             Yerevan~36, Armenia}
\address[LZ-Sol]{Soltan Institute for Nuclear Studies,
             90950~Lodz, Poland}
	     
\ifx AA
\makeatletter
\begingroup
  \global\newcount\c@sv@footnote
  \global\c@sv@footnote=\c@footnote     
  \output@glob@notes  
  \global\c@footnote=\c@sv@footnote     
  \global\t@glob@notes={}
\endgroup
\makeatother
\fi

\newpage

\begin{abstract}
The attenuation of the electron shower size beyond the shower maximum
is studied with the KASCADE extensive air shower experiment
in the primary energy range of about $10^{14}-10^{16}$~eV.
Attenuation and absorption lengths are determined 
by applying different approaches, including the method of
constant intensity, the decrease of the flux of extensive air
showers with increasing zenith angle, and its variation with
ground pressure.  We observe a significant dependence of the
results on the applied method.  The determined values of the
attenuation length ranges from 175 to 196 g/cm$^2$ and of the
absorption length from 100 to 120 g/cm$^2$.  The origin of these
differences is discussed emphasizing the influence of intrinsic
shower fluctuations.
\end{abstract}


\end{frontmatter}


\section{Introduction}

The longitudinal development of the electron shower size $N_e$ in extensive
air showers (EAS) is characterized by an approximately
exponential decline for atmospheric depths well beyond the shower
maximum.  Therefore, $N_e$ shows a strong
dependence on the slant depth.
The magnitude of this effect is usually described by
two different quantities, the attenuation and the absorption
length.

\Markline{-65mm}{65mm}
{ \Mark In the following the attenuation length 
$\Lambda_{N_e}$ describes the average
decrease of the electron number $N_{e}$ with increasing 
atmospheric depth $X$ in showers selected to be similar
in primary energy:
\begin{equation}
\label{equ:att}
\langle N_e(X)\rangle \propto \exp{(-X/\Lambda_{N_e})}
\end{equation}

The
absorption length $\Lambda_{rate}$ is defined to parameterize
the decrease of the integral flux
$j(>N_e)$ of showers with electron numbers greater than $N_e$ 
at a given atmospheric depth $X$:
\begin{equation}
\label{equ:abso}
j(> N_e,X) \propto \exp{(-X/\Lambda_{rate}})
\end{equation}
}

Both quantities are frequently used to rescale showers to a
certain angle of incidence \cite{yoshida95,ave01}, i.e.\
atmospheric depth, or for applying ground pressure corrections. 

There exist various different methods to determine the attenuation 
respectively the absorption length.
The purpose of this paper is to
determine $\Lambda_{N_e}$ and $\Lambda_{rate}$ 
from one set of experimental data by applying different
methods.  The results are then compared to each other in order
to estimate systematic uncertainties resulting from the different
approaches.

It should be emphasized that there is a conceptual difference
between the attenuation and the absorption length.  The
absorption length $\Lambda_{rate}$ depends on the form and steepness of the
primary energy spectrum and is only defined for a set of showers
while the definition of the attenuation length $\Lambda_{N_e}$
is independent of that and applicable to individual showers.
\Markline{-20mm}{25mm}
{\Mark However, the methods used in this analysis do not allow to
measure the attenuation length $\Lambda_{N_e}$ on a shower-by-shower
basis.
Therefore only an effective $\Lambda_{N_e}$ is measured which
describes the shape of a mean shower at atmospheric depths
much larger than the depth of the shower maximum.}

Assuming a pure composition and a
power law of the electron shower size spectrum, the relation
between attenuation and
absorption length reads $\Lambda_{N_e}=(\gamma-1)\cdot
\Lambda_{rate}$, with the spectral index $\gamma$.  For a mixed
composition or other spectral shapes, the relation is much more
complicated.
A further difficulty in these relations are shower fluctuations
which depend on shower size and angle and modify the shower
size spectra.

The definition of the absorption length $\Lambda_{rate}$ in Equation (\ref{equ:abso})
is different
from the one frequently used to infer proton-air cross section
from EAS with the constant $N_e$-$N_{\mu}$-Method (e.g.~\cite{hara83} and
Chapter 6).
In the following, absorption of showers without any preselection by
the muon shower size is analyzed.

The paper is organized in the following way.  Chapter 2 describes
the experimental setup and the reconstruction of 
EAS with the KASCADE experiment.
Two different methods to determine the attenuation length
are presented in Chapter 3.  Both are based on a selection of
showers above approximately the same primary energy.  This is done
by considering the shift of the knee position with
increasing zenith angle or by applying constant intensity cuts to
spectra in different angular bins.  Chapter 4 describes the
analysis of the spectra with regard to the absorption length by
examining the variation of the rate with angle of incidence
or atmospheric ground pressure.  Since all quantities related to
EAS, especially the electron shower size, are influenced by 
intrinsic shower fluctuations, we estimate the magnitude of these
effects in Chapter 5.
A short remark about the already mentioned constant $N_e$-$N_{\mu}$-Method
is given in Chapter 6
and a summary is given in Chapter 7.

\section{KASCADE - experimental setup and EAS reconstruction}

KASCADE (KA{\em rlsruhe} S{\em hower} C{\em ore and} A{\em rray}
DE{\em tector}) \cite{doll,antoni02} is located at Forsch\-ungs\-zentrum Karlsruhe,
Germany ($8.4^{\mathrm{o}}$~E, $49.1^{\mathrm{o}}$~N) 
at \mbox{110 m} a.s.l.
corresponding to an average vertical atmospheric depth of \mbox{1022 g/cm$^2$}.
The experiment measures the electromagnetic, muonic, and hadronic
components of EAS with three major detector systems, a large field
array, a muon tracking detector, and a central detector.

\begin{figure}[!bt]
\begin{minipage}[b]{.46\textwidth}
\centering\epsfig{file=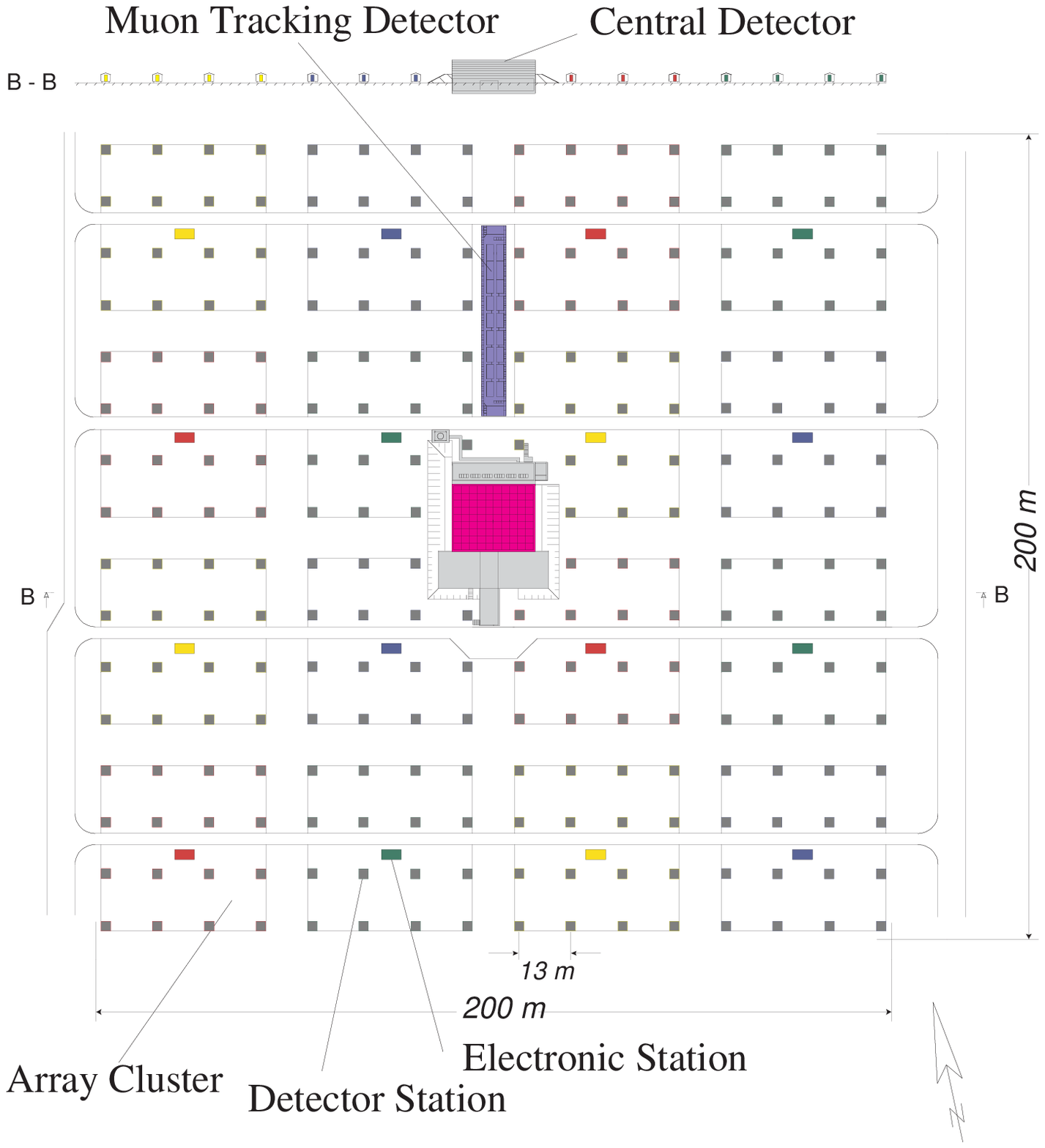,width=\linewidth}
\end{minipage} \hfill
\begin{minipage}[b]{.50\textwidth}
\centering\epsfig{file=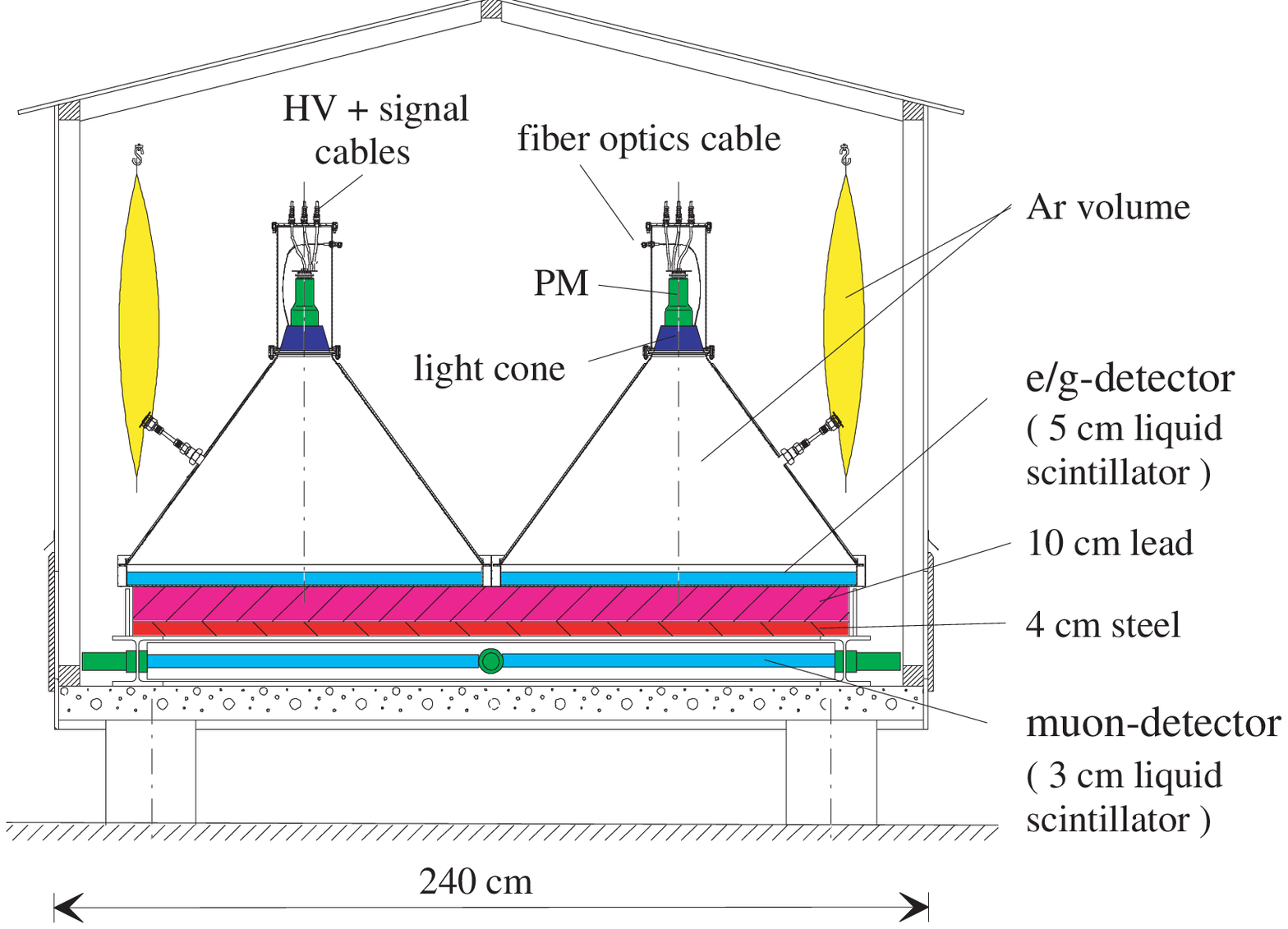,width=\linewidth}
\end{minipage}
\caption{\label{fig:array} Schematic view of the KASCADE field array 
and a detector station}
\end{figure}

In the present analysis data from the 200 $\times$ 200 m$^2$ scintillation 
detector array shown in Figure \ref{fig:array} are used.
The 252 detector stations of the array are uniformly spaced on a grid 
with a separation of 13~m. The stations are organized in 16 electronically
independent clusters with 16 stations in the 12 outer and 15 stations in the
four inner clusters.
The array stations contain four detectors measuring the electromagnetic component
in the four inner resp.~two
in the 12 outer clusters. They are filled with \mbox{5 cm} thick custom made liquid
scintillator with an area of 0.8 m$^2$ each.
The 192 stations in the 12 outer clusters are additionally equipped
with four sheets of plastic scintillation detectors, \mbox{90 $\times$ 90 $\times$ 3 cm$^3$} each.
These detectors, mounted below 10 cm absorber of lead and 4 cm of iron, which
corresponds to more than 20 radiation lengths and to a muon energy threshold of 
\mbox{230 MeV}, are used to measure the muonic component of an air shower.
The energy sum, the arrival time of the first shower particle in a station, 
and the hit pattern of the detectors are read out individually for the muon
and e/$\gamma$ detectors.
The trigger conditions are a detector multiplicity of 10/20 out of 32/60
e/$\gamma$ detectors fired in at least one of the outer/inner clusters.

The scintillator array reaches full trigger and reconstruction efficiency for showers with
log$_{10}N_e>4.3$ almost independent of the type of the primary particle.
In the following analysis, only showers safely above this threshold
are used.

\par


In an iterative shower reconstruction procedure the
electron shower size and slope of the lateral electron distribution
('age') are determined
 mainly by maximizing a likelihood function describing
the measurements with the
Nishimura Kamata Greisen (NKG) formula~\cite{nk,g} assuming a
Moli\`ere radius of 89~m. 
Detector signals are corrected for contributions of other particles,
i.e. the electromagnetic detectors for contributions of muons,
gammas, and hadrons.
A detailed presentation of these corrections can be found
in \cite{antoni}.

Shower cores are identified in a first step by the center of
gravity of the energy deposits in the detectors and in a second
iteration by the NKG-fit mentioned above.  Shower directions are
determined by fitting a conical shower front to the measured
particle arrival times.

For the present analysis showers have been selected with 
reconstructed shower cores inside a circle of 91~m around the center 
of the array and with zenith angles less than 40$^{\mathrm{o}}$.
This results in a set of about 37 million well reconstructed showers.

The influence of the sampling fluctuation on the results of the following
analysis is examined in Chapter \ref{ch:mc}.
Detailed descriptions of the detectors and reconstruction
performances are given in \cite{doll,antoni02}.

\section{Attenuation length ${\mathbf \Lambda_{{\mathbf N_{\mathbf e}}}}$}
\label{cha:atten}

The idea of the following analysis is to select showers of a certain
primary energies and then to observe the attenuation of their electron shower
size $N_e$ with increasing
atmospheric depth.  Assuming a functional dependence between the primary
energy spectra and the electron shower size spectra, application of equal
intensity cuts to the integral electron shower size spectra in different angular
bins select showers above approximately equal primary energies.  This is
the method of constant intensity \cite{nagano}.

Assuming the knee to be of astrophysical origin, i.e.\ to be a
feature of the primary energy spectrum, the method of constant
intensity can be applied in a special way to determine the
attenuation of the shower size at the knee.  One has to keep in
mind, however, that the shapes of electron shower size spectra are
influenced by intrinsic shower fluctuations which in turn show
variations with electron shower size and incident angle of the
EAS.
Therefore, the knee positions
in the electron shower size spectra 
is shifted in a non-trivial way and differently in the various bins of zenith angle
towards larger shower sizes.

Attenuation is observed in the following by analyzing showers in different 
angular bins. Equation (\ref{equ:att}) then becomes:
\begin{equation}
\label{equ:secatt}
\langle N_e(\Theta )\rangle \propto \exp{(-{X_0 \over \Lambda_{N_e}} (\sec \Theta -1) )} .
\end{equation}
$X_0=1022$ g/cm$^2$ is the average vertical atmospheric depth at the KASCADE
observation level and $\Theta$ is the zenith angle of the EAS.

\subsection{Position of the knee}

\begin{figure}
\centering\epsfig{file=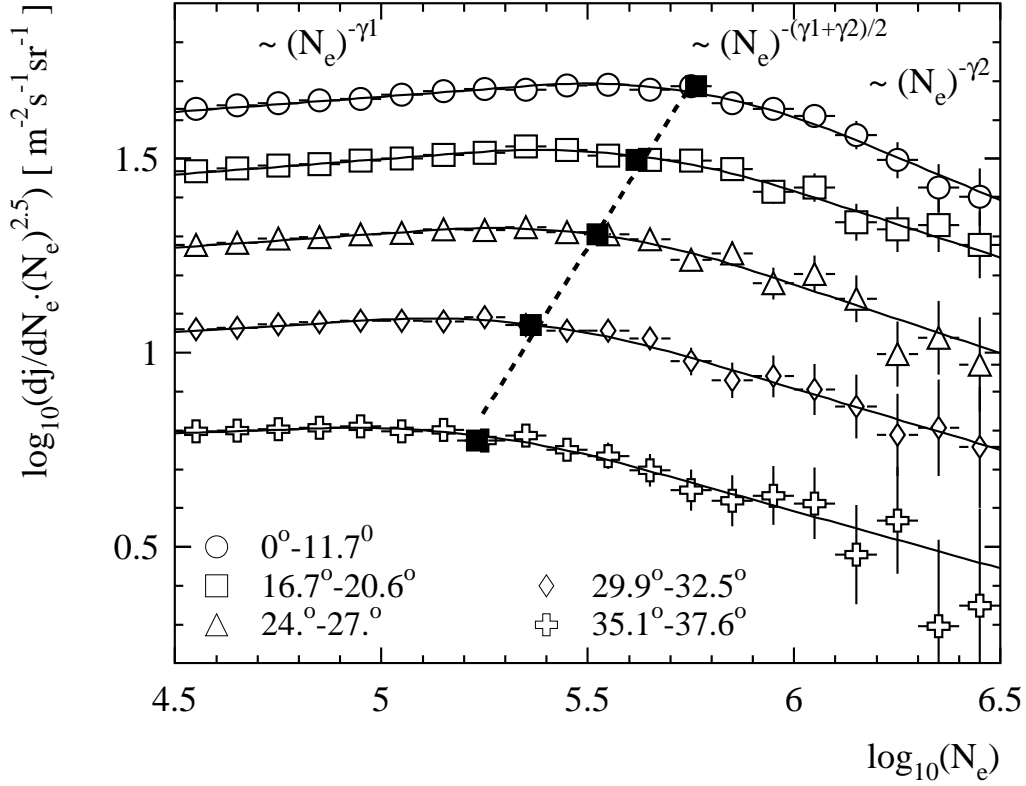,width=\linewidth}
\caption{\label{fig:kniespek} Differential 
electron shower size spectra measured by the KASCADE experiment.
The shift of the knee position with 
increasing zenith angle is indicated by the dotted line.
The solid lines indicate the fits described in the text.
Only size spectra of every second angular bin are plotted.}
\end{figure}

Figure \ref{fig:kniespek} shows the differential electron shower size
spectra in the knee region as measured with the KASCADE
experiment for different angular bins.  The shift of the
knee position with increasing atmospheric depth is clearly
visible.

Spectra in 10 different angular bins ranging from
$0^{\mathrm{o}}$ to $40^{\mathrm{o}}$ were analyzed.
The angular bins are selected to cover the same solid angle.
For reasons
of clarity, only every second spectrum is plotted in
Figure \ref{fig:kniespek}.  The spectra are fitted with the following function
\cite{ralph_durb}:

\begin{eqnarray}
\label{equ:knee}
dj/dN_e \propto \left\{ {\begin{array}{ll}
N_e^{-\gamma_1} & \log_{10}{N_e} \le \log_{10}{N_{e,k}}-\epsilon  \\
N_e^{a_2\log_{10}{N_e+a_1}} & \log_{10}{N_{e,k}}-\epsilon < 
\log_{10}{N_e} < \log_{10}{N_{e,k}}+\epsilon\\
N_e^{-\gamma_2} & \log_{10}{N_e} \ge \log_{10}{N_{e,k}}+ \epsilon  \\
\end{array}}
\right.
\end{eqnarray}
with the electron shower size at the knee $N_{e,k}$.  Fit
parameters are the two spectral indices, $\gamma_{1}$,
$\gamma_{2}$, the size at the knee, $\log_{10}{N_{e,k}}$, the half width $\epsilon$,
and the
differential flux at the knee.  The half width $\epsilon$ of the knee
region is constant within the statistical errors and therefore
fixed to $\epsilon=0.38$ on a logarithmic shower size
scale in these fits.
The coefficients $a_2$ and $a_1$ are determined by the condition
that the function and its first derivative are continuous at the
boundary of the knee region.
The knee position is defined as
the point where the local spectral index of the function above is equal to
the mean value of the slopes $\gamma_{1}$ and $\gamma_{2}$. Some 
results of the fits are summarized in Table \ref{tab:knee}.

\begin{table}
\caption{\label{tab:knee} Results of the fits of Equation (\ref{equ:knee}) to the electron
size spectra. 
}
\begin{center}
\begin{tabular}{rrc}
\multicolumn{3}{c}{} \\[0.25ex]
\hline
\multicolumn{1}{c}{\raisebox{-0.5ex}{angular bin}} 
&\multicolumn{1}{c}{\raisebox{-0.5ex}{position}} 
&\multicolumn{1}{c}{\raisebox{-0.5ex}{integral flux}} \\
& \multicolumn{1}{c}{\raisebox{+0.5ex}{of the knee}} &
\multicolumn{1}{c}{\raisebox{+0.5ex}{above the knee}} \\
& \multicolumn{1}{c}{\raisebox{+1.5ex}{$\log_{10}(N_e)$}} &
\multicolumn{1}{c}{\raisebox{+1.5ex}{$[10^{-8}\ {\mathrm m}^{-2}\ {\mathrm s}^{-1}\ {\mathrm {sr}}^{-1}]$}} \\
\hline
\rule{0mm}{3.5mm}$0.0^{\mathrm{o}}-11.7^{\mathrm{o}}$ &  
$5.76 \pm 0.03$ & $6.47 \pm 0.35$ \\
$ 11.7^{\mathrm{o}}-16.7^{\mathrm{o}}$ & 
$5.65\pm 0.03$ & $8.43 \pm 0.40$ \\
$ 16.7^{\mathrm{o}}-20.6^{\mathrm{o}}$ &
$5.62\pm 0.03$ & $7.58\pm 0.36$ \\
$ 20.6^{\mathrm{o}}-24.0^{\mathrm{o}}$ &
$5.59\pm 0.03$ & $6.73\pm 0.38$ \\
$ 24.0^{\mathrm{o}}-27.0^{\mathrm{o}}$ &
$5.52\pm 0.03$ & $6.57 \pm 0.43 $ \\
$ 27.0^{\mathrm{o}}-29.9^{\mathrm{o}}$ &
$5.46\pm 0.04$ & $6.49 \pm 0.48$ \\
$ 29.9^{\mathrm{o}}-32.5^{\mathrm{o}}$ &
$5.36\pm 0.04$ & $6.80 \pm 0.49$ \\
$ 32.5^{\mathrm{o}}-35.1^{\mathrm{o}}$ &
$5.33\pm 0.04$ & $5.41 \pm 0.44$ \\
$ 35.1^{\mathrm{o}}-37.6^{\mathrm{o}}$ &
$5.23\pm 0.05$ & $5.56 \pm 0.45$ \\
$ 37.6^{\mathrm{o}}-40.0^{\mathrm{o}}$ &
$5.07\pm 0.06$ & $6.70 \pm 0.64$ \\
\hline
\end{tabular}
\end{center}
\end{table}

\begin{figure}
\centering\epsfig{file=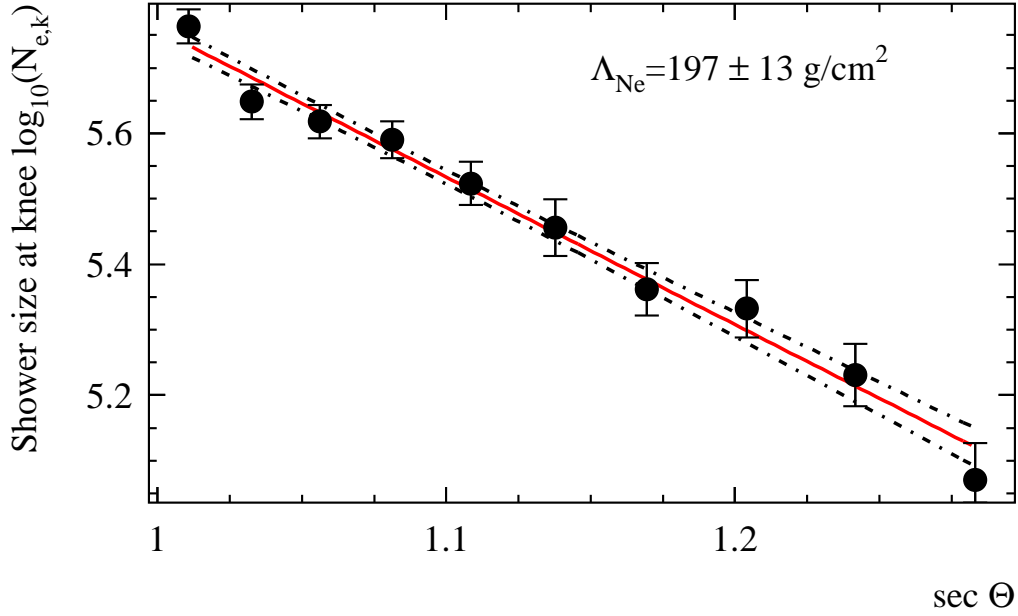,width=\linewidth}
\caption{\label{fig:knieab} Knee positions in the differential 
electron shower size spectra for different angular bins. The lines show an 
exponential fit and the error region of the fit.}
\end{figure}

The positions of the individual knees are plotted in Figure \ref{fig:knieab}.
The knee positions are shifted exponentially with increasing
atmospheric depth.
A fit according to Equation (\ref{equ:secatt}) has been applied.
The attenuation length determined from these
values is $\Lambda_{N_e}=197\pm 13$ g/cm$^2$.  The quoted error is of
statistical kind and follows from the errors of the fit.  The
resulting attenuation length has to be taken with some precaution. 
Using other fit functions to determine the knee position yield
different results.  For example, defining the knee
position as the intersection of two straight lines on a logarithmic
scale results in an attenuation length of $170 \pm 36$ g/cm$^2$.  In
Ref.~\cite{chili}, an attenuation length of \mbox{$222\pm28$ g/cm$^2$} for
KASCADE data is reported using yet another definition of the knee
position. All these results are compatible within their 
statistical errors, but their relatively large spread indicates the uncertainty
resulting from the choice
of the function describing the shower size spectra.

Attenuation lengths of other experiments determined by
the attenuation of the shower size at the knee are e.g. for
EAS-TOP 222$\pm$3 g/cm$^2$ at an atmospheric depth of 820
g/cm$^2$ \cite{eastop} and for MAKET ANI 302$\pm$71 g/cm$^2$ at
700 g/cm$^2$ \cite{chili}.
The tendency of increasing values with decreasing atmospheric depths
reflects the change in the longitudinal shower profile while
approaching the shower maximum.
A comparison of these results with those from KASCADE is therefore
not straightforward.

A constant integral flux above the knee would be a verification of
the assumptions made in the method of constant intensity.  The
last column in Table \ref{tab:knee} shows the values for the
integral fluxes above the knee.  A confirmation of the results of
the EAS-TOP experiment of a constant flux above the knee within an
error of 20\% \cite{eastop} is not possible with the numbers
obtained here.  Our results match better e.g.~with a hypothesis of 
decreasing flux with increasing zenith angle which could be
motivated by the influence of intrinsic shower fluctuations. 
Again, it must be noticed here that the integral fluxes depend
on the method applied to determine the knee positions. 
The errors of the integral fluxes above the knee quoted in Table
\ref{tab:knee} are of statistical kind only. The systematic
errors due to the choice of the fit function are by far larger. 
This makes any interpretation concerning a constant flux above the
knee very difficult.

\subsection{Method of Constant Intensity}
\label{ch:mci}

\begin{figure}
\centering\epsfig{file=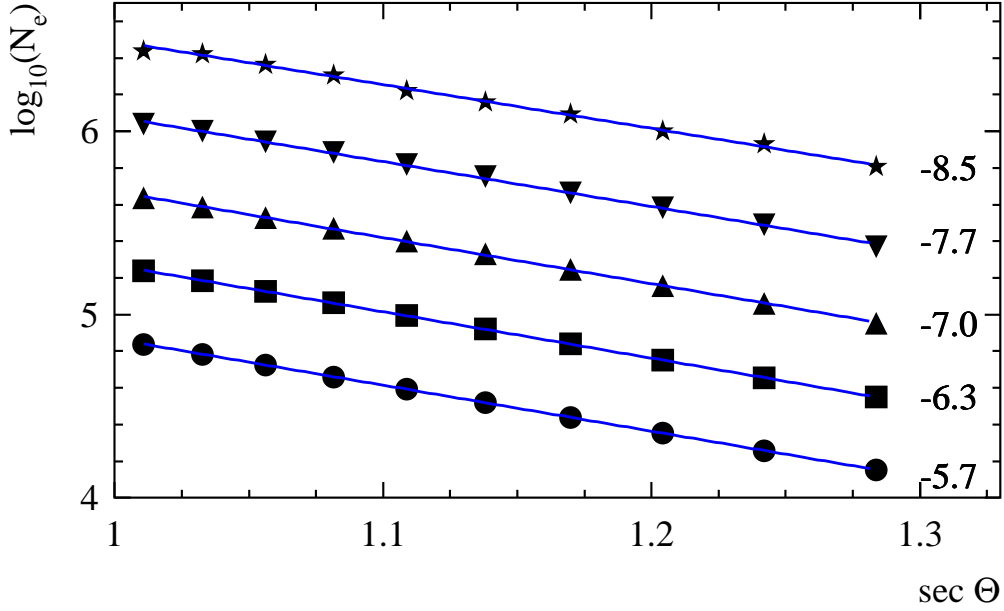,width=\linewidth}
\caption{\label{fig:secne} Electron number as a function of
zenith angle for different values of constant intensity.  Error
bars are smaller than the point sizes.  The lines indicate fits
using exponential functions, the numbers on the right side are the
respective logarithmic fluxes $\log_{10}{(j(>N_e))}$ in
m$^{-2}$s$^{-1}$sr$^{-1}$.}
\end{figure}

\begin{figure}
\centering\epsfig{file=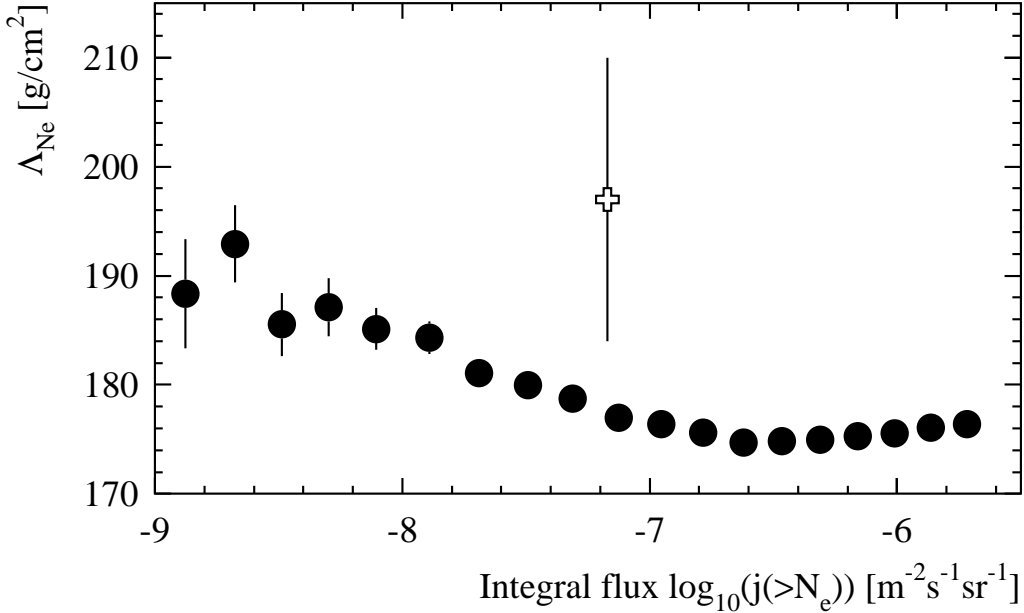,width=\linewidth}
\caption{\label{fig:mci} Attenuation lengths determined by the
method of constant intensity (filled points).  The attenuation
length determined by the knee position is represented by the open
symbol, plotted at the weighted mean flux at the knee of 
\mbox{$j(>N_{e,k} ) = ( 6.75 \pm 0.13 ) \cdot 10^{-8}$  m$^{-2}$s$^{-1}$sr$^{-1}$}.}
\end{figure}

To infer the attenuation length by the method of constant
intensity the integral electron shower size spectra were grouped into
ten bins of zenith angle ranging from 0$^{\mathrm{o}}$ to
40$^{\mathrm{o}}$.  The $N_{e}$ shifts of the spectra were then
analyzed for 18 different values of integral intensities ranging from $6
\cdot 10^{-5}$ to $5 \cdot 10^{-8}$ m$^{-2}$s$^{-1}$sr$^{-1}$. 
Figure \ref{fig:secne} shows the observed attenuation of the
shower size with increasing angle for five selected flux values.
Again, an exponential behavior is visible. 
The results of exponential fits to the electron shower sizes at
different zenith angles for constant intensities are shown in
Figure \ref{fig:mci}.  Each point in this figure corresponds to a
certain intensity, which increases from left to right, i.e. the
primary energy of the EAS increases from right to left.

From this analysis attenuation lengths between 174 and 190
g/cm$^2$ are obtained.  The result shows an increase of
$\Lambda_{N_e}$ with decreasing flux.  This effect is partly
due to the correlation between primary energy and atmospheric
depth of the shower maximum.  Measurements closer to the shower
maximum show a flattening of the longitudinal shower profile which
is related to increasing values of attenuation lengths.  The weak
increase of $\Lambda_{N_e}$ observed towards higher fluxes is in
parts due to increasing statistical reconstruction uncertainties
for smaller showers.  Monte Carlo simulations, described in 
Chapter \ref{ch:mc}, confirm this expectation. However, the simulations 
indicate a stronger effect (5-10 g/cm$^2$) than observed in the data. 
This difference is most likely due to the simplified nature of these 
simulations.

The difference between the attenuation length obtained from
the decrease of the shower size at the knee with increasing depth
compared to those determined by the method of constant
intensities reflects the methodical uncertainties.  The more or
less arbitrary definition of the position of the knee and the
strong influence of intrinsic shower fluctuations (see Chapter \ref{ch:mc})
to both methods are reasons for the observed deviations.

\section{Absorption length ${\mathbf \Lambda_{\mathbf {rate}}}$}
\label{ch:abs}

The absorption length is determined from the integral electron shower
size spectra by analyzing the decrease of the shower rate with
increasing atmospheric depth for constant $N_e$.  In
the following, the atmospheric depth is changed both by the
zenith angle of the showers and by using data taken at different
atmospheric ground pressures.
The main difference between the two methods is the range of variations
in slant depth, with the angular method it is up to 300 g/cm$^2$ and it is about
50 g/cm$^2$ only with the barometric method.
Varying the shower zenith angle in the barometric method,
a precise determination of the absorption length can be
obtained for different $N_e$ and at different effective
atmospheric depths.
For example flux variations of showers with $\log_{10}{(N_e)}>5.4$
as function of depth are shown in Figure \ref{fig:baroab1}. 

As mentioned in the introduction, the value of the absorption
length is usually related to the attenuation length by
multiplication with the local spectral index of the integral
electron shower size spectrum.  However, this implies the assumption of
single component composition and a pure power law spectrum, both
of which are not correct.  Since the factors
in the relation are sensitive especially to the large change in
the spectral index in the range of the knee, we refrain from
applying this relation to calculate $\Lambda_{rate}$ from
$\Lambda_{N_e}$.

\subsection{Angular Method}

\begin{figure}
\centering\epsfig{file=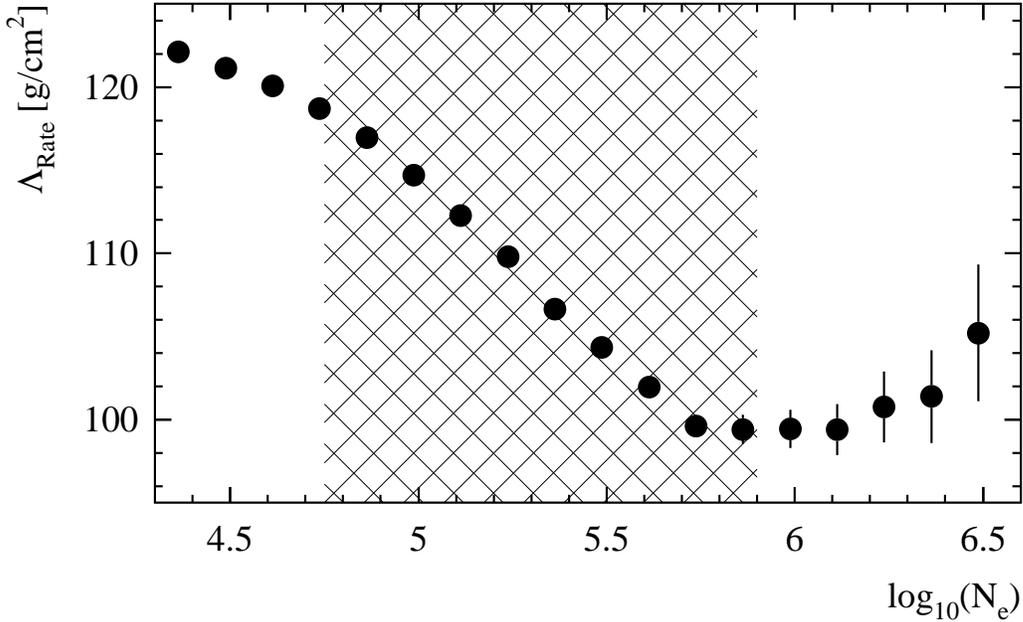,width=\linewidth}
\caption{\label{fig:absoabso} Absorption lengths versus electron shower size 
determined by the angular method.
The hatched area indicates the region
influenced by
the changing slopes of the spectra in the knee region.}
\end{figure}

Using the angular method, the absorption length is determined
from the integral electron shower size spectra by analyzing the decrease
of the flux with zenith angle for given electron shower sizes.  An
exponential fit of the decreasing flux according to Equation
(\ref{equ:abso})
then yields the absorption length.  Results are shown in
Figure \ref{fig:absoabso} for different thresholds in electron
shower size. 

The overall trend of decreasing absorption lengths with
increasing electron number is caused by the knee in the
electron shower size spectra.  Scanning from low to high values of $N_e$
over the knee
region with its changing slopes at different electron shower sizes in
different angular bins mimics a stronger reduction of the fluxes,
thereby resulting in a decline of $\Lambda_{rate}$.
The region where the angular method is not applicable  due to this bias 
is indicated by the hatched area in
Figure \ref{fig:absoabso}.
However, the
observed small increase of $\Lambda_{rate}$ at shower sizes above 
$10^{6.1}$ electrons, well beyond the knee, exhibits the same 
behavior as the attenuation lengths in Figure \ref{fig:mci}
and is related to an effective decrease of the distance
between observation level and shower maximum.

\subsection{Barometric Method}

\begin{figure}
\centering\epsfig{file=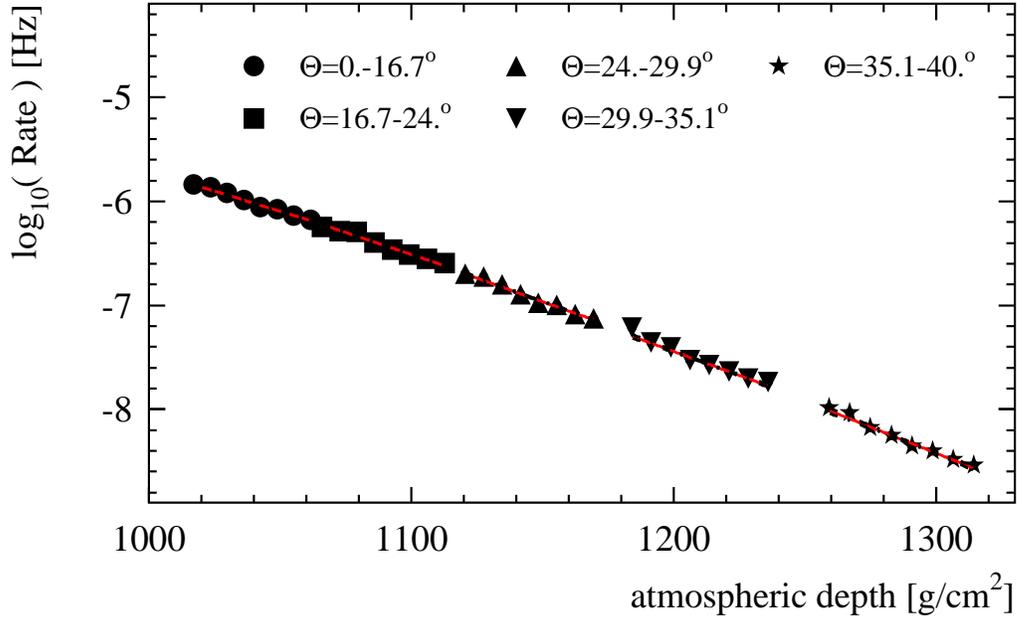,width=\linewidth}
\caption{\label{fig:baroab1} Variation of showers with atmospheric 
ground pressure and zenith angle for showers with $\log_{10}{N_e}>5.4$.
The lines indicate fits with exponential functions. } 
\end{figure}

The influence of atmospheric ground pressure is analyzed by counting
the number of showers above a certain electron shower size in two hour time
intervals.  The variation of ground pressure within these time
intervals is negligible.
A change in the ground
pressure of 1 hPa results in a variation of the rate by about 0.7-1\%
depending on the zenith angle.
 The mean values of the rate depending on
ground pressure and zenith angle, i.e. atmospheric depth are plotted
for showers with $\log_{10}{(N_e)}>5.4$
in Figure \ref{fig:baroab1}.
{\Mark
The ground pressure variation is converted to 
atmospheric depth by 
\begin{equation}
X(p(t)) = \frac{X_0}{\cos{\bar{\Theta}}} + \frac{p(t)-p_0}{g\cdot\cos{\bar{\Theta}}}.
\end{equation}
\Markline{-25mm}{50mm}

$X_0 = 1022$ g/cm$^2$ is the average vertical atmospheric depth at the KASCADE
observation level,
$\bar{\Theta}$ the mean zenith angle of the angular bin, $p(t)$ the time
dependent 
atmospheric
ground pressure, $p_0$ the 
mean atmospheric ground pressure and $g$ the gravitational acceleration .
}

\begin{figure}
\centering\epsfig{file=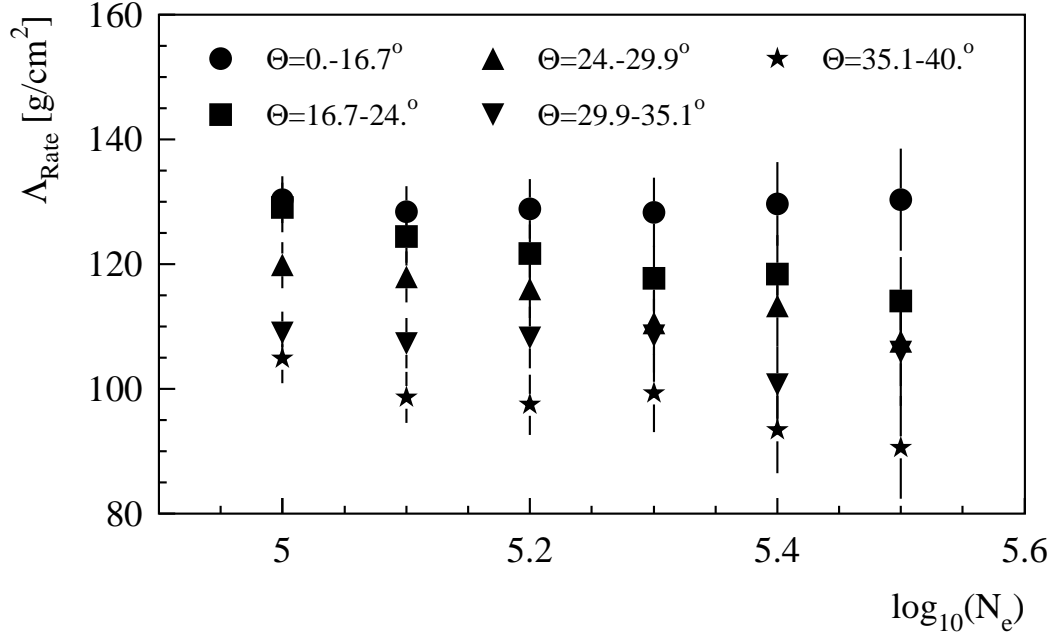,width=\linewidth}
\caption{\label{fig:barone1} Absorption lengths determined with the
barometric method for showers with different zenith angles.}
\end{figure}

Again, the decrease of intensity is fitted by the exponential law
of Equation (\ref{equ:abso}).
These fits are performed for six different
thresholds in electron shower size and for five different angular bins. 
Figure \ref{fig:barone1} shows the absorption lengths obtained by
these fits.  Both figures, \ref{fig:baroab1} and \ref{fig:barone1} show 
a decreasing absorption length with
atmospheric depth.
This means a deviation of the longitudinal development from the 
simple exponential form of Equation (\ref{equ:abso}).
Simulated EAS obtained by CORSIKA calculations show a similar
behavior.

The absorption lengths determined that way decrease
from 130 g/cm$^2$ in the first angular bin to 90-100 g/cm$^2$ in the
last one.  The results of the angular and the barometric method
are consistent with a decrease of the absorption length by about
10 g/cm$^2$ in the observed size region due to the already mentioned
change in slope of the spectra.

\section{Influence of intrinsic shower fluctuations and 
reconstruction accuracy on the method of constant intensity}
\label{ch:mc}

As mentioned before, intrinsic shower fluctuations, 
reconstruction accuracy of the shower size $N_e$,
and detector response influence the
results of most analyses of EAS including
the method of constant intensity.  To get an
estimate of this, a computational analysis using a
parameterization of fully simulated EAS is
performed.

Electron shower size spectra are simulated with
the intrinsic shower fluctuations 
and the detector response switched on or off.
More precisely, the shower size spectra are obtained from a hypothetical
primary energy spectrum by
\begin{equation}
\label{equ:int}
{dj \over d\log_{10} N_e} = \sum_{i=p,fe} \int_{-\infty}^{\infty} 
{dj_i \over d \log_{10} E} \ p^i(\log_{10} N_e | \log_{10} E ) d\log_{10} E
\end{equation}

\par
\begin{figure}[!bt]
\begin{minipage}[b]{.50\linewidth}
\centering\epsfig{file=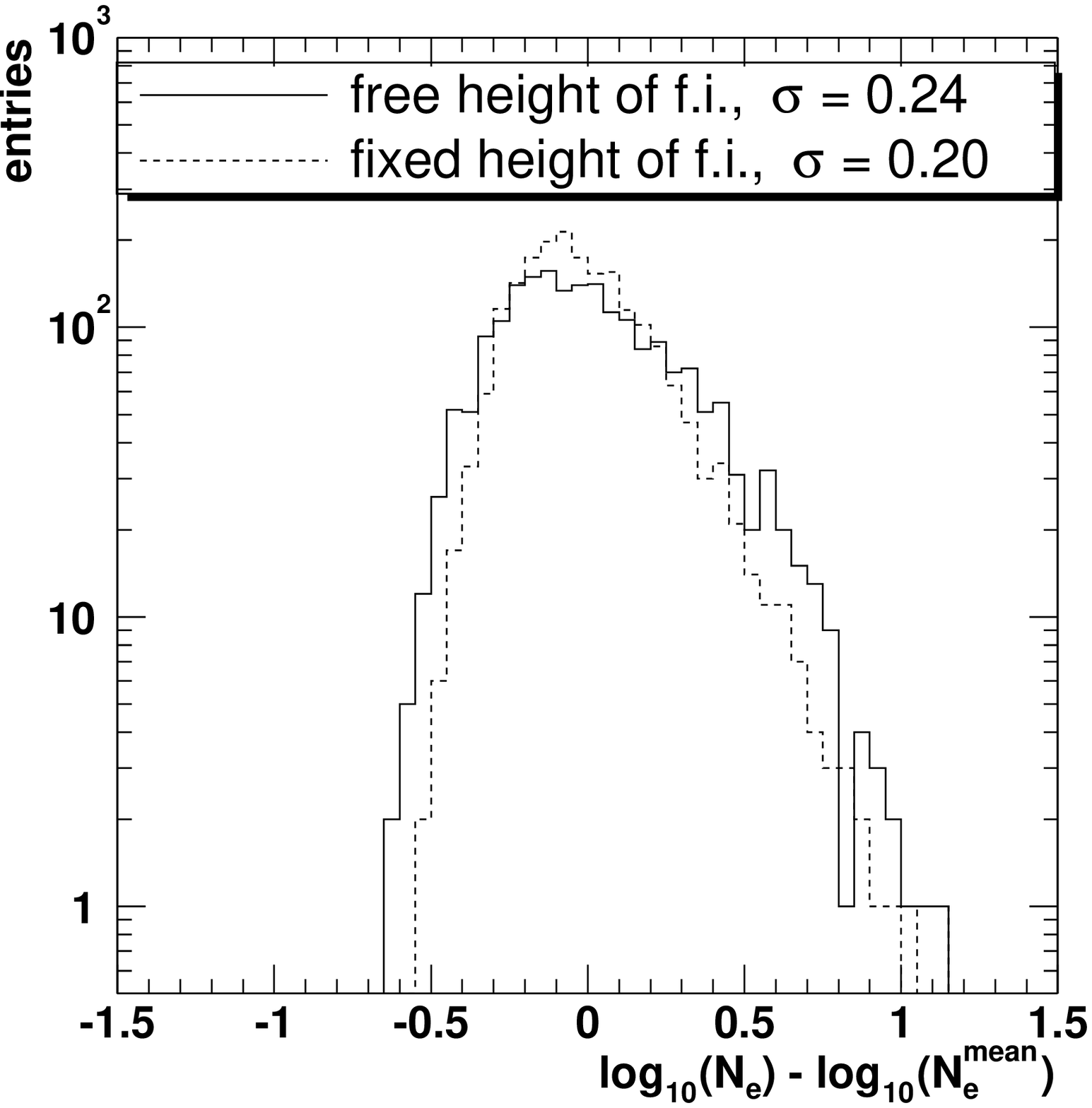,width=\linewidth}
\end{minipage} \hfill
\begin{minipage}[b]{.50\linewidth}
\centering\epsfig{file=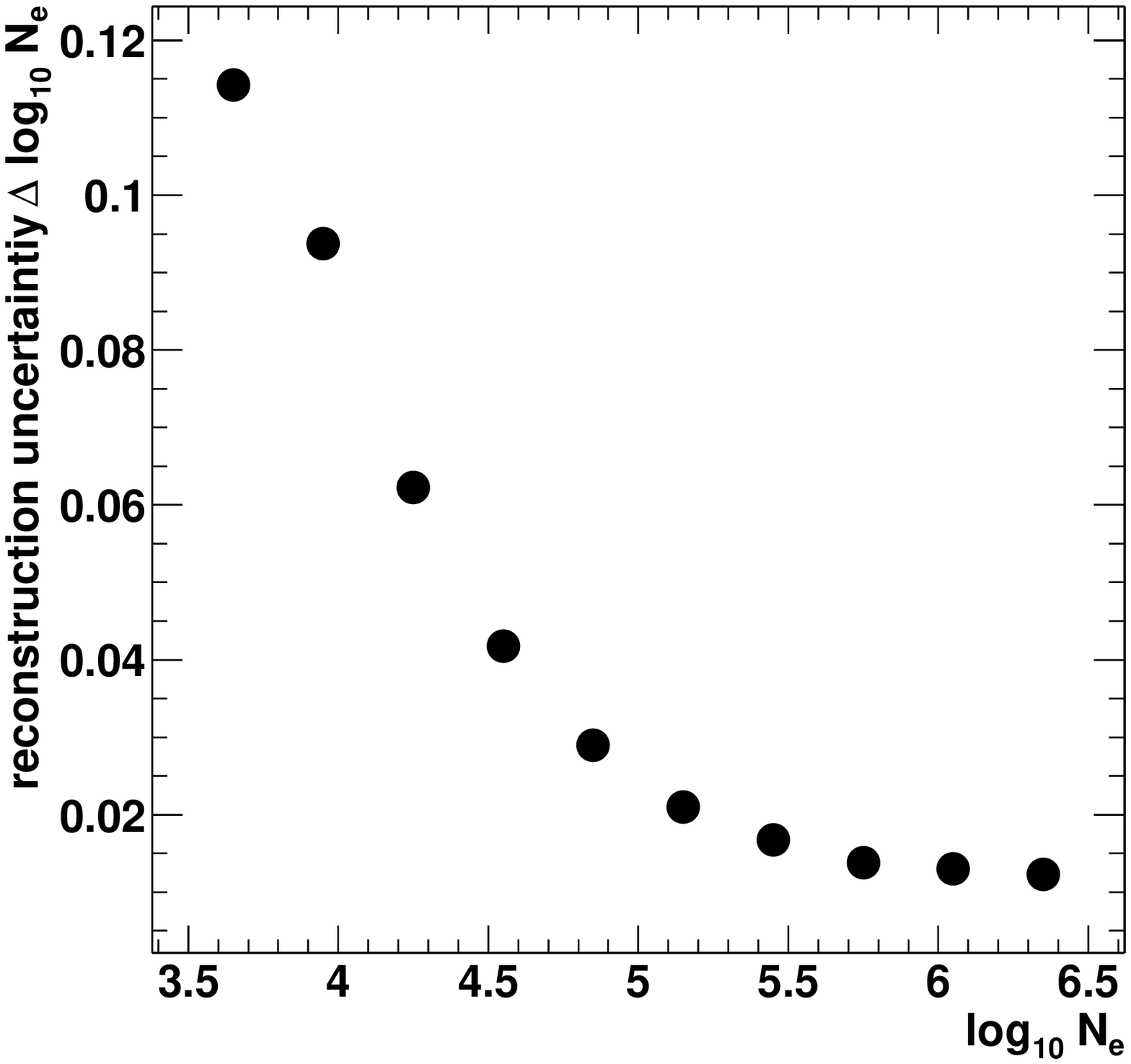,width=\linewidth}
\end{minipage}
\caption{\label{fig:fluk}
Left: Intrinsic shower fluctuation for 1 PeV proton showers and 22 degrees zenith
angle
determined with CORSIKA simulations (solid line). 
The dashed histogram shows the fluctuations for 1 PeV proton showers with
fixed first interaction (f.i.) height of 23.9 km.
The distributions are shown relative to their mean value of $\log_{10}{N_e} \approx 4.6$.
Right:
reconstruction uncertainty from CORSIKA simulations including detector
simulations.
}
\end{figure}

The integral is evaluated by simulation of the primary energy spectra
and calculation of the corresponding $N_e$ values, followed by 
a smearing of the $N_e$ values according to the reconstruction uncertainties
and shower fluctuations.
The primary energy
spectrum is  based on an analysis of 
electron and muon size spectra \cite{ralph}.
which exhibits a knee
at 4 PeV with spectral indices of $-2.7$ below and $-3.1$ above the
knee.  The composition is represented by a light (proton) and a
heavy (iron) component where the protons account for the knee in
the total energy spectrum.

\Markline{-35mm}{30mm}
{\Mark The dependency of the electron shower size on the zenith angle is
accomplished by using Equation (\ref{equ:att}) with a constant 
attenuation length of $\Lambda_{N_e} = 150$ g/cm$^2$.
This means that all simulated showers have the same longitudinal 
development.
The different longitudinal profiles in actually measured showers
alter the shower size spectra additionally to the effects considered
in this chapter but are not taken into account in this analysis.
}

The kernel function $p^i$ factorizes into three parts, the intrinsic
shower fluctuations $s^i$, the efficiency $\epsilon^i$ for the 
triggering of the measurements, and the reconstruction accuracy $r^i$ 
of the electron shower size $N_e$.
\Markline{-30mm}{40mm}
{\Mark All three functions are determined with
air shower simulations using
the CORSIKA \cite{corsika} package (Version 5.635) with QGSJET
\cite{qgs} as hadronic interaction model and EGS4 \cite{egs} for
the electromagnetic part of the shower.
The subsequent detector simulations are based on the GEANT \cite{geant} package
and all simulated showers are reconstructed by the same procedure as
measured data.}

For the fluctuations $s^i$ and the reconstruction accuracy $r^i$
shower size dependent parameterizations of the mean values
of Gaussian fit functions to the distributions of the
detailed shower simulations
are used in the following Monte Carlo calculations.
Figure \ref{fig:fluk} (left) shows as an example one of the distributions
used in these parameterizations.
The primary particles in the Figure \ref{fig:fluk} (left) are protons with
fixed energies of 1 PeV and 22 degrees zenith angle.
Mean and width of this distribution are 4.64 and 0.24 on a logarithmic scale
in electron shower size.
The right figure shows the statistical uncertainty of the reconstruction
versus the electron shower size.
As can be seen, the reconstruction error
is only important at small electron shower sizes below about $\log_{10}(N_e) < 4.5$.

\begin{figure}
\centering\epsfig{file=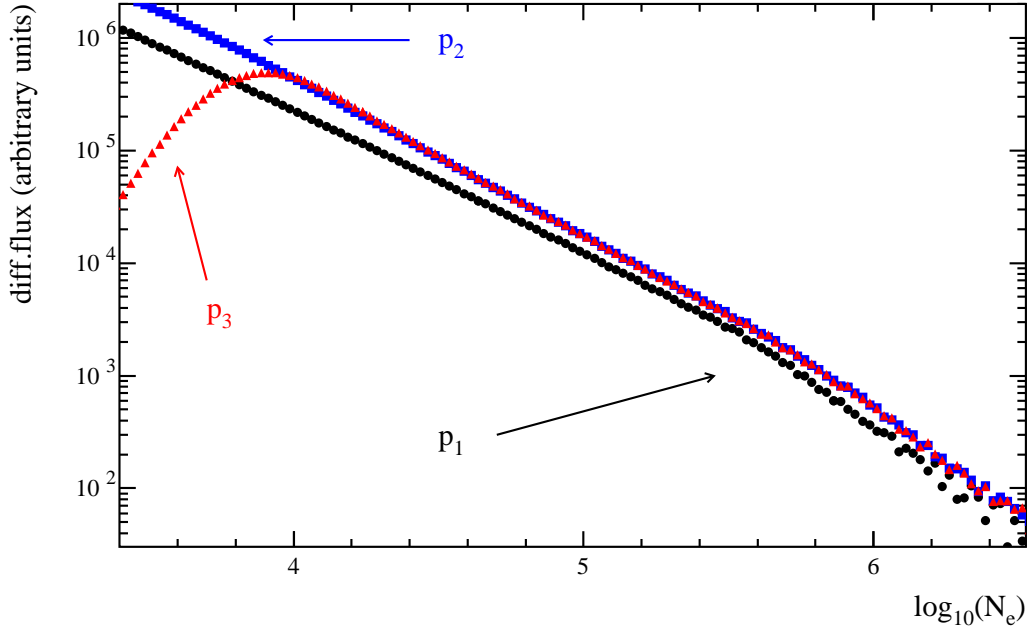,width=\linewidth}
\caption{\label{fig:newu} Monte Carlo shower size spectra with
different assumption for the kernel function $p^i$ of Equation 
(\ref{equ:int}).
$p_1$ is the spectrum obtained without any fluctuations, $p_2$ includes
intrinsic shower fluctuations and $p_3$ additionally the reconstruction accuracy
as well as the trigger efficiencies.
}
\end{figure}

Three different sets of electron shower size spectra are obtained by calculating
the integral in Equation (\ref{equ:int}) numerically by Monte Carlo methods.
The first one ($p^i_1$), without any fluctuations and detector effects, i.e.
$p^i_1 = \delta( E - f(N_e))$, the second one ($p^i_2$) with only shower fluctuations
on, and the third one \mbox{( $p^i_3$ )} with the full kernel function as obtained by shower
and detector simulations described above.
Each set
consists of size spectra in the same angular bins as in
the data analysis presented in Chapter \ref{cha:atten}.
Figure \ref{fig:newu} shows three of these Monte Carlo spectra for all
different assumptions on $p^i$.
The intrinsic shower fluctuations shift the spectra to higher values of
both 
electron shower size and intensity and change the slope of the spectra.
The reconstruction accuracy has only little influence while the
trigger and reconstruction efficiencies becomes more and more important
below $\log_{10}(N_e) < 4.5$.

\begin{figure}
\centering\epsfig{file=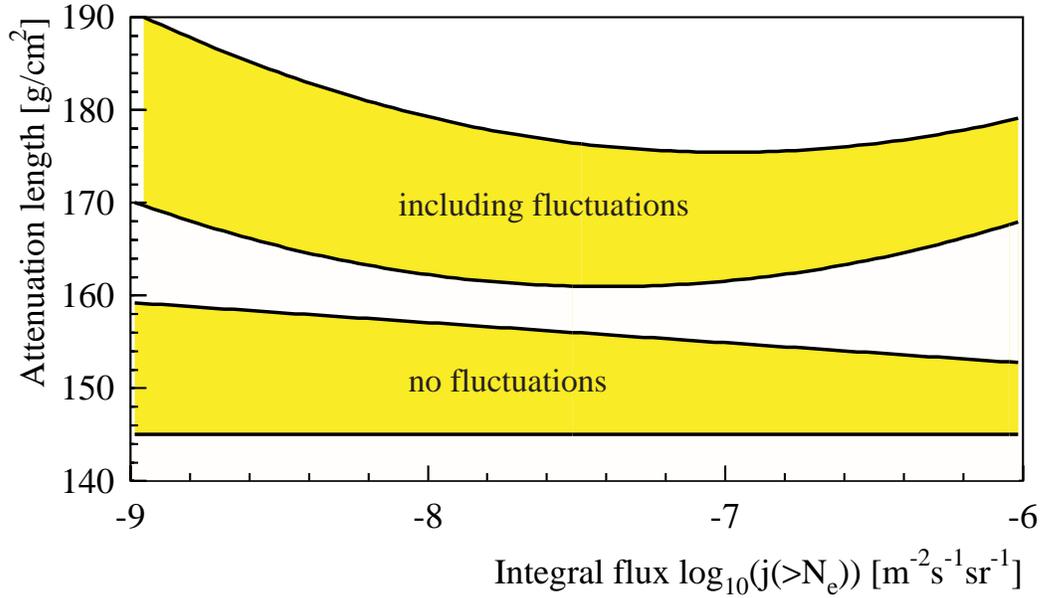,width=\linewidth}
\caption{\label{fig:mcmci} Attenuation lengths from Monte Carlo
spectra determined with the method of constant intensity.  The
shaded regions reflect the uncertainties due to the choice of the
input parameters of the MC. The attenuation lengths in the lower
region are determined from average shower size spectra (kernel function
$p^i_1$), the upper
ones from taking intrinsic shower fluctuations, statistical
reconstruction errors and detector efficiency (kernel function
$p^i_3$) into account.}
\end{figure}

Applying the method of constant intensity to all sets yields the
bands of attenuation lengths plotted in Figure \ref{fig:mcmci}.
For reasons of clarity only the results for $p^i_1$ and $p^i_3 $ 
are plotted.
The shaded regions reflect the estimated range of the uncertainty due
to the choice of the parameters of the primary energy spectra, composition, and
the parameterization of the information from CORSIKA and detector
simulations. 

The results of Figure \ref{fig:mcmci} demonstrate that fluctuations
strongly influence the results of the method of constant
intensity; the attenuation lengths are shifted towards higher
values by 15-30 g/cm$^2$.  The attenuation lengths obtained from
the mean conversion of the primary energy spectrum to size
spectra with kernel function $p^i_1$ reflect the assumed constant
$\Lambda_{N_e} \approx 150 $ g/cm$^2$ and are in fact quite
different from the experimental results shown in
Figure \ref{fig:mci}.  On the other hand, the results obtained
from simulations taking detector response and
intrinsic shower fluctuation into account, i.e. kernel function $p^i_3$, 
resemble the data very much.  This
shows that shower fluctuations have a major influence on the
results of the previous analysis in Chapter \ref{ch:mci}.  Any
interpretation of the attenuation lengths has to take this effect
into account.
As all the other methods presented in this
paper are also sensitive to the form of the shower size spectra,
this conclusion applies there as well.

The Monte Carlo study provides also some information about the
dependence of the attenuation length on the fluxes.  The increase
towards higher fluxes is due to the increasing statistical
reconstruction uncertainty for smaller shower sizes.  In the low
flux region, results are influenced by the changing shape of the
spectra due to the knee and due to intrinsic shower fluctuations. 
The very strong increase of the attenuation length with
decreasing fluxes in Figure \ref{fig:mci} is not fully reproduced
in this solution of Equation (\ref{equ:int}).
One reason for that is the changing slope of the longitudinal
shower profile while getting closer to the shower maximum, i.e. here
with increasing electron shower size.
This effect is not included in the simulations.
Another reason could be the
simplified parameterization of the fluctuations by Gaussian
functions in $\log_{10}{N_e}$.  Simulations show especially for the proton component
asymmetric distributions with very long tails towards larger
electron shower sizes (see Figure \ref{fig:fluk}).  Also the assumption of
a two component composition instead of a more realistic one is
a strong simplification and influences these results.
However, the presented sensitivity of the method of constant intensity
to shower fluctuations and the trend of observing attenuation
lengths with a bias towards larger values are not affected by
these simplifications.

\section{Remarks about absorption lengths and p-air inelastic cross sections}

The term absorption length is often used in literature in connection
with p-air inelastic cross sections.
This absorption length, in the following called $\Lambda_{cross}$,
describes something different than the one
used in this paper.

The idea of the cross section analysis, the so-called constant $N_e$-$N_{\mu}$ 
method is based on the assumption
that EAS with about the same $N_e$ and $N_{\mu}$ at ground level have developed
through the same slant depth
\cite{hara83}. Analyzing the zenith angle dependence of the
flux of these
shower samples determines then
$\Lambda_{cross}$.
The samples are enriched of proton initiated showers by selecting showers
with the largest number of electrons.
The proton-air cross section is then calculated by the formula
\mbox{$\Lambda_{cross} = k \cdot 14.5 \cdot m_p / \sigma^{inel}_{p-Air}$}.
The simple factor $k$ is used to correct for the influence of
shower fluctuation and detector response \cite{hara83,honda,aglietta99}.
It includes properties of the underlying hadronic interaction model.

{\Mark A good correlation of the position of the first interaction
point to the electron shower size at observation level is required
for the \mbox{$N_e$-$N_{\mu}$} method.
\Markline{-70mm}{80mm}
For example Figure \ref{fig:fluk} shows the $N_e$ distribution of
1 PeV proton showers.
The dotted line is the result of simulations using a fixed depth
of the first interaction.
In case of no intrinsic shower fluctuations (i.e. perfect correlation
between depth of first interaction and observed electron shower size)
this dotted curve would be a narrow peak at $\langle \log_{10}{N_e} \rangle 
\approx 4.6$.
The comparison with the result of the full simulations shows that
intrinsic shower fluctuations dominate over fluctuations due to
the first interaction point.
Similar conclusions, but for higher energies, were reached in an 
independant study in \cite{alvarez02}.
This means that a shower with large observed $N_e$ is not necessarily
a shower
starting deep in the atmosphere (see Fig.7 in \cite{alvarez02}).
Therefore, it appears doubtful whether the constant \mbox{$N_e$-$N_{\mu}$}
method could give
reliable results concerning the proton air-cross section
for primary energies relevant for KASCADE.}

\section{Summary}

We have presented a comprehensive measurement of the atmospheric
attenuation of the electromagnetic component in EAS. The data are
based on the detector array of the KASCADE experiment.  Both the
attenuation and absorption length have been reconstructed for shower
sizes in the interval $\log_{10} N_e=4.5$ to $\log_{10}N_e=6.5$
covering the energy range of the knee.  The
most important experimental findings can be summarized as:

\begin{itemize}
\item The attenuation length increases with electron numbers
$N_e>10^{5.5}$ from $\Lambda_{N_e}=175$ g/cm$^2$ to 194 g/cm$^2$. 
\item The angular method for the determination of the absorption
length appears almost 
inapplicable in the knee region
due to the changing slopes in the size spectra.
Well above the knee values between
100 and 110 g/cm$^2$ were obtained.
\item The longitudinal development deviates from a simple exponential
assumption and steepens with increasing atmospheric depth.
\item Intrinsic shower fluctuations cause an increase of the
attenuation lengths by about 15-30 g/cm$^2$.  This shift depends
on the shower size and cannot be neglected.
\end{itemize}


Furthermore, the experimental studies emphasize the need to
account for methodical effects as well as for shower and sampling
fluctuations when attempting quantitative interpretations of the
attenuation and absorption lengths.
It is hoped that the
detailed analyses of experimental and simulated data presented
here will help to improve the understanding of the experimental
input to calculations, thereby reducing systematic
differences between different approaches.

\section*{Acknowledgments}
\vspace*{-8mm} The authors are indebted to the members of the
engineering and technical staff of the KASCADE collaboration who
contributed with enthusiasm and engagement to the success of the
experiment.  The work has been supported by the Ministry for
Research and Education of the Federal Government of Germany. 
The collaborating institute in Lodz is supported by the Polish
State Committee for Scientific Research (grant No. 5 P03B 133 20)
and the Institute of
Bucharest by a grant of the Romanian National Academy for
Science, Research and Technology.  The KASCADE work is embedded
in the frame of scientific-technical cooperation (WTZ) projects
between Germany and Romania, Poland, and Armenia.

\end{document}